\documentclass[a4paper,11pt]{article}
\usepackage{pos}
\newcommand{\gama}{$\gamma$}
\newcommand{\fermi}{\textsl{Fermi}}

\newcommand{\apj}{ApJ}
\newcommand{\mnras}{MNRAS}
\newcommand{\aap}{A\&A}
\newcommand{\nat}{Nature}

\title{Exploring the variability properties of gamma-ray emission from blazars}
 \ShortTitle{$\gamma$-ray variability of blazars}

  \author*[a]{Gopal Bhatta}

\author[b]{Niraj Dhital}

\affiliation[a]{Institute
of Nuclear Physics Polish Academy of Sciences,\\
  PL-31342 Krakow, Poland}
  
  \affiliation[b]{Central Department of Physics, Tribhuvan University\\
 Kirtipur 44613, Nepal}

\emailAdd{gopal.bhatta@ifj.edu.pl}
\emailAdd{niraj.dhital@cdp.tu.edu.np}

\abstract{We present  results of variability study of a sample of 20 powerful blazars using Fermi/LAT (0.1--300 GeV) observations.  We studied decade-long observations applying various analysis tools such as flux distribution, symmetry analysis, and RMS-flux relation. It was found that the $\gamma$-ray flux distribution closely resembles a log-normal probability distribution function and can be characterized by linear RMS-flux relation. The power spectral density analysis  shows the statistical variability properties of the sources as studied are consistent with flicker noise, an indication of long-memory processes at work. Statistical analysis of the distribution of flux rise and decay rates in the light curves of the sources, aimed at distinguishing between particle acceleration and energy-dissipation timescales, counter-intuitively suggests that both kinds of rates follow a similar distribution and the derived mean variability timescales are on the order of a few weeks. The corresponding emission region size is used to constrain the location of $\gamma$-ray production sites in the sources to be a few parsecs. Additionally, using Lomb-Scargle periodogram and weighted wavelet z-transform methods and extensive Monte Carlo simulations, we detected year-timescale quasi-periodic oscillations in the sources S5 0716+714, Mrk 421, ON +325, PKS 1424-418, and PKS 2155-304. We also performed recurrence quantification analysis of the sources and directly measure the deterministic quantities, which suggest that the dynamical processes in blazars could be a combination of deterministic and stochastic processes, while some of the source light curves revealed significant deterministic content.}

\FullConference{37$^{\rm{th}}$ International Cosmic Ray Conference (ICRC 2021)\\
		July 12th -- 23rd, 2021\\
		Online -- Berlin, Germany}

\begin{document}
\maketitle

\section{Introduction}
The \gama-ray sky is mainly dominated by the emission from the discrete extra-galactic sources known as blazars. 
Blazars are the most luminous sources in the Universe that mostly shine in the non-thermal emission. The sources can be identified with some of the extreme features such as large luminosity, rapidly variable broadband (from radio to TeV emission) flux and polarization. These extreme properties of blazars are largely ascribed to the Doppler boosted emission from the kpc/Mpc scale relativistic jets, which transport matter and momentum into the inter-galactic medium. The sources receive their large luminosity as the matter from the pc scale accretion disk gradually swirls inwards before being engulfed by the supermassive black hole, lurking at the center. The jet is powered via the Blandford-Znajek  process \cite{Blandford1977}, in which the rapidly spinning black hole immersed in the magnetic field of the accretion disk twists the magnetic field into helical structure creating electric field required launch the jet.

A commonly adopted scheme for classification of blazars is based on the features in their spectra. More luminous sources with emission lines over the continuum are referred to as flat-spectrum radio quasars (FSRQ), whereas less luminous ones showing weak or no such lines are referred to as BL Lacertae (BL Lac). Also, in case of FSRQs, the synchrotron peak lies in the lower frequency region and it is generally believed that the external Compton process is responsible for the high energy emission, as opposed to synchrotron self-Compton process \citep[][]{Ghisellini1998}. This is corroborated by the fact that these sources are known to have abundant seed photons from accretion disk, broad-line region and dusty torus \citep{Ghisellini2011}. On the other hand, the synchotron peak lies in the higher frequency region (optical or X-rays) for the BL Lac objects. This class includes extreme sources and feature high energy emission from few tens of keV to TeV energies believed to be produced by synchrotron and inverse Compton processes. Apparent low luminosity of these sources are possibly due to the absence of strong circumnuclear photon fields and relatively low accretion rates. Blazars can also be classified on the basis of the frequency of synchrotron peak ($\nu_\mathrm{s}$) into three categories-- the high synchrotron peaked blazars (HSP; $\nu_s > 10^{15}$ Hz), the intermediate synchrotron  peaked blazars (ISP; $10^{14}<\nu_s <10^{15} $ Hz), and  the low synchrotron peaked blazars (LSP; $\nu_s < 10^{14}$ Hz) \cite{Fan2016}. It is observed that the bolometric luminosity of blazars and the synchrotron peak frequencies are related. According to the blazar sequence, a scheme used to explain this observation, both the bolometric luminosity and the $\gamma$-ray emission decrease in the direction from FSRQs to HSP, but the peak frequencies move towards higher energies \citep{Fossati1998,Ghisellini2017}. In addition, FSRQs are mostly $\gamma$-ray dominant, whereas in HSP blazars, synchrotron and $\gamma$-ray emission are comparable.

In leptonic models, the radiation output is largely contributed by electrons--positron  pairs, and   although
the jets might contain protons in abundance, owing to their much larger mass, the particles
do not produce significant radiative output. Nevertheless, they can transport a significant  amount of the momentum and kinetic power of the jet (see, e.g., \cite{Sikora2000}). In the hadronic scenario, the protons can be accelerated up to the ultra-relativistic energies such that the \gama-ray emission is mainly produced via synchrotron radiation from protons (e.g., [19,20]) or through photo-pion production (e.g., [21,22]). The neutral pions subsequently 
decays into photons, or in case of charged pions, into pairs and neutrinos. These ultra-relativistic secondary electrons/positrons lose their energy quickly due
to synchrotron radiation. Both these synchrotron photons and the initial $\pi^0$ decay photons have too
high energy to escape $\gamma-\gamma$ absorption in the source, thus initiating synchrotron-supported pair cascade.
(e.g., [23]). Moreover, following Hillas Energy relation, the size of the tentative size of an accelerator that is capable of accelerating particles up to tens of EeV is of the order of Kpc--Mpc. As the Mpc scale radio jets provide most favorable conditions to the acceleration of the non-thermal particles to the highest energies,  blazar jets could be promising avenues to the production of UHECR. The capability of the current instruments e.g. Fermi/LAT to precisely measure photon fluxes and the spectra from a wide range of astrophysical objects has sparked interest in using the \gama-ray sky as a probe for various hadronic interaction processes that are linked with the \gama-ray emission. Therefore, in such case, \gama-ray emission can serve as a probe to the UHECR. Moreover, it compliments the studies that put to the test theories about the origin of ultrahigh-energy cosmic rays (UHECRs).

Blazars are usually variable in practically all accessible bands, on diverse timescales and
amplitudes. The general  nature of the variability appears to be aperiodic, with the exception of a few sources such as Mrk 501, Mrk 421, S5 0716+714, PKS 1424 which are reported to show quasi-periodic oscillations \citep[][]{Bhatta2019,Bhatta2020a}. The sources in general are far away in the sense that the the telescopes with current capability hardly can resolve the central engine. In such context, Study of variability serves as one of the most effective tool to probe the innermost structures, emission mechanism: the rapid variability in the timescale of a few hours can probe probe the innermost structures, the flaring episode that last weeks to month provided insight into the particle acceleration and cooling mechanisms taking during the propagation of shock waves or rampant magnetic re-connection events; whereas periodic variability could be sign of the presence of binary supermassive black holes. 

In this proceeding, we present some of the highlights of the work on \gama-ray variability study of a sample of 20 blazars using a decade-long observation from Fermi/LAT telescope (see \cite{Bhatta2020a,Bhatta2020b}. In Section 2, the analysis methods are discussed and in Section 3 result and discussion are presented. Finally in the last section, a summary of the study is followed.

\begin{table*}
        \caption{General information about the sample Fermi/LAT blazars}
        \centering
        \label{table:1}
        \begin{tabular}{l|l|l|l|c}
                \hline
                Source name &Redshift (z) &Source class&FV& $\beta$\\
                \hline
	Mrk 421 	&	0.03	&	BL Lac 	&	43.65$\pm$1.45	&	1.00	$\pm$0.08	\\
	Mrk 501 	&	0.0334	&	 BL Lac 	&	33.47$\pm$	3.76	&	 1.10$\pm$10	\\
	1ES 1959+65 	&	0.048	&	BL Lac 	&	49.55$\pm$2.84	&	1.10	$\pm$0.14	\\
	BL Lac 	&	0.068	&	BL Lac 	&	64.10$\pm$1.05	&	1.0$\pm0.10$ 	\\
	W Comae 	&	0.102	&	BL Lac 	&	24.70	$\pm$8.87	&	1.10	$\pm$0.09	\\
	PKS 2155-304 	&	0.116	&	BL Lac 	&	45.93	$\pm$2.02	&	0.90$\pm$0.20 	\\
	ON +325 	&	0.131	&	BL Lac 	&	43.78	$\pm$4.60 	&	0.80	$\pm$0.14	\\
	3C 273 	&	0.158	&	FSRQ 	&	94.66$\pm$0.98 	&	1.20$\pm$0.17 	\\
	S5 0716+714 	&	0.3	&	BL Lac 	&	62.20	$\pm$1.05 	&	1.00	$\pm$0.15 	\\
	4C +21.35 	&	0.432	&	FSRQ 	&	114.91$\pm$0.59 	&	1.10$\pm$0.12	\\
	3C 66A 	&	0.444	&	BL Lac 	&	58.43	$\pm$1.78 	&	0.90	$\pm$0.17	\\			
	3C 279 	&	0.536	&	FSRQ 	&	104.29$\pm$	0.46	&	1.10$\pm$0.16	\\
	 TON 0599 	&	0.7247	&	BL Lac 	&	111.69$\pm$0.88	&	1.30	$\pm$0.15 	\\
	3C 454.3  	&	0.859	&	FSRQ 	&$81.30\pm0.30$		& 1.30$\pm0.17$		\\
	AO 0235+164 	&	0.94	&	BL Lac 	&	95.53	$\pm$1.12 	&	1.40	$\pm$0.19	\\
	PKS 0454-234 	&	1.003	&	BL Lac 	&	68.25	$\pm$1.06	&	1.10	$\pm$0.09 	\\
	CTA 102 	&	1.037	&	FSRQ 	&	117.42	$\pm$0.37 	&	1.20$\pm$0.19	\\
	PKS 1424-418 	&	1.522	&	 FSRQ	&	70.44$\pm$0.69	&	 1.5$\pm$0.13	\\
	4C+38.41 	&	1.813	&	FSRQ 	&	92.99	$\pm$0.72 	&	1.2$\pm$0.15 	\\
	PKS 1502+106 	&	1.84	&	FSRQ 	&	90.11	$\pm$0.70	&	1.3$\pm$0.10 	\\
                \hline
        \end{tabular}
\end{table*}

\section{Observations and data processing}
\label{sec:2}

The \gama-ray observations of the sample sources used in this study were made by the Large Area Telescope (LAT) onboard Fermi Gamma-ray Space Telescope \cite{Atwood2009}.
The LAT detector has a large effective area ($> 8000\ cm^{2}$) which collects high energy photons incident on it from a wide field-of-view ($>$ 2.4 sr, or about 20\% of the whole sky). It has a high angular resolution which is about $3.5^\circ$ around 100 MeV reaching less than $0.15^\circ$ at energies above 10 GeV. We processed Pass 8 data from the \fermi/LAT 3FGL catalog using Fermi Science Tools{\footnote{\url{https://fermi.gsfc.nasa.gov/ssc/data/analysis/software/}}} to obtain light curves of the sample sources following the standard procedures of the unbinned likelihood analysis\footnote{\url{https://fermi.gsfc.nasa.gov/ssc/data/analysis/scitools/likelihood_tutorial.html}}. We selected the photon-like events in the energy range 100 MeV -- 300 GeV, using the criteria $evclass=128$ and $evtype=3$ and incident upon a circular region of interest (ROI) of $10^{\circ}$ radius centered around the source. To eliminate the Earth limb events,  zenith angle was limited to  $ <$ 90$^{\circ}$.
For the analysis, Galactic diffuse emission model and isotropic model for point sources were used alongside the \fermi/LAT 3FGL catalog, whereas for the diffuse source response, the Galactic and extra-galactic models of the diffuse $\gamma$-ray emission, namely, \emph{ gll\_ iem v06.fit} and \emph{ iso\_P8R2 SOURCE V6 v06.txt} were used. To construct the  blazar \gama-ray light curves, Fermi/LAT observations were binned in weekly bins and  the task \emph{ gtlike} was run to carry out maximum-likelihood analysis, and  the events with test statistic 10 (equivalently $\gtrsim 3 \sigma$) \citep{Mattox1996} were applied to extract the source flux. For details of the data processing readers are directed to \cite{Bhatta2020a}.

\section{Results and Discussion}
\label{sec:3}
\subsection{Time series analysis,}
The  variability observed the the sample light curves were quantified by computing their fractional variability (FV; see \cite{Vaughan2003}) usually written as
\begin{equation}
F_{var}=\sqrt{\frac{S^{2}-\left \langle \sigma _{err}^{2} \right \rangle}{\left \langle F \right \rangle^{2}}},
\end{equation}
where $\left \langle F \right \rangle$, S$^2$, and $ \left \langle \sigma _{err}^{2} \right \rangle$ represent the mean flux, variance and mean error square, respectively. The FV values for the sample blazar are list in the 4th column of Table \ref{table:1}. It can be seen that the sources exhibit remarkable variability, as high as 100\%, in the \gama-ray band. This provides a strong impetus to study the physical process driving the it.

The \gama-ray in blazars could be produced at both the innermost of the region of the central engine, where most of the gravitational potential energy get processed in to electromagnetic radiation, creating a strong seed photon field required for the Inverse-Compton scattering to result in the observed \gama-ray emission. In such scenario, the observed variability could result owing to the modulations in a number of factors, such as magnetic field, viscosity, instantaneous accretion rate, etc. Moreover, the \gama-ray variable emission can be ascribed to the strong shock and/or magnetic re-connection events,  that take place at the kilo-parsec-scale radio jets aligned within $\sim10^{o}$ to the line of sight, near the region from where the radio knots are seen being ejected.  Such events inject high energy particles which consequently can up-scatter the nearby photon or can instigate pion decay process.

\subsection{Flux distribution and  RMS-flux relation} 
 
The histogram of the \gama-flux were  were fitted normal and log-normal PDF. Between the two log-normal PDF  was found to be a better representation of the blazar flux distribution. Moreover, a linear trend of RMS-flux relation was evident all the sample sources.  The observed log-normal distribution of the blazar flux suggests multiplicative coupling of these perturbations at the disk, as opposed to additive processes. 
The observed log-normal distribution of the blazar flux can be modeled in terms of disk processes.  The fluctuations  in the disk, contributing to flux variability, can take place at different radii and thereby be dictated by viscosity fluctuations in accordance with the local viscous timescales. In turn, these modulate the mass accretion rates at larger distances from black hole. Variable emission from accretion disks owing to variable accretion rate could be driven by uncorrelated fluctuations in the $\alpha$-parameter taking place at different radii of the disk \citep[see][]{Lyubarskii1997}. 
In the context of relativistic jets, such distribution could be a natural consequence of emission from Poynting flux dominated jets that hosts mini-jets distributed isotropically within the emission region and that get ejected close to the line of sight with a high bulk Lorentz factor $\sim 50$.  In  such scenario,  the resulting flux distribution has been found to hold the RMS-flux relation \citep[see][]{Biteau2012}.

 \subsection{Power spectral density} 
The variability power distribution of variable sources over temporal frequencies can be studied using the discrete Fourier periodogram (DFP) of their light curves. For an evenly sampled time series $x(t_{j})$ sampled at times $t_{j}$ where $j = 1,2,..,n$, and spanning a total duration of observation $T$, the DFP at a temporal frequency $\nu$ is expressed as
\begin{equation}
P\!\left(\nu \right)=\frac{1}{n} \, \left | \sum_{j=1}^{n}x\!\left( t_{j} \right ) \, e^{-i 2\pi \nu t_{j}} \right |^{2} \, .
\end{equation}
The frequency range between frequencies $\nu_{min}=1/T$ and $\nu_{max}=1/2 \Delta t$, where $\Delta t$ is mean sampling step in the light curve, is evenly sampled for $n/2$ frequencies in log-space and the periodograms are computed for these frequencies. Normalizing it with a factor $2 T/\left(n \bar{x} \right)^{2}$ so that the unit becomes $(rms/mean)^{2} /d$,  one can obtain the total integrated power of periodogram which is nearly equal to the variance of the light curve. The power-law PSD was fitted to the periodogram using several slope indexes using the method PSRESP, a MC simulation method that accounts for the artifact due to real astronomical observations. The best fit slope indexes of the power-law for the sample sources, as listed in the 5th column of Table 1, centers around unity. Such power-law variability is famously known as \emph{flicker noise} and frequently observed in the nature (e. g. see \cite{Press1978}).

The power-law type PSD observed in most cases can be explained both in the context of a turbulent flow behind propagating shock \citep[][]{Marscher1992} or a standing/reconfinement shock in blazar jets \citep{Marscher2008}. Emission from a single dominant turbulent cell may get enhanced due to Doppler boosting, contributing to the temporal frequency corresponding to the size or velocity of the cell. Due to the stochastic nature of the turbulence, cells of various sizes and thus various frequencies will be Doppler enhanced over time depending on their velocities and angles of sight. Consequently, this will give rise to a variability spectrum  over wide range of frequencies that is in agreement with the power-law noise seen in blazars. Also, in case the magnetic field at the accretion disk is fairly magnetized due to the material accreted over reasonable period of time, the magnetic field can extract vast amount of rotational energy by threading the black hole and channel the energy into the jet as the bulk power of the relativistic jets. As the radiation power is only about 10\% of the total jet power, a significant contribution 
 to the jet contents could be provided by poynting flux \citep{Ghisellini2014}. This can then facilitate the rampant magnetic reconnection events triggering stochastic particle acceleration and energy dissipation at various temporal and spatial scales. If the observed variable \gama-ray emission is  produced in such a scenario, the variability power spectrum should closely resemble  power-law shape.

 \subsection{Quasi-periodic oscillations}
 One of the widely used methods for periodicity search in the astronomical time series, is the Lomb-Scargle periodogram (LSP; \cite{ Lomb76,Scargle82}), The method approached the unevenly sampled observations and containing gaps  by  least-square fitting sine waves of the form $X_{f}(t)= A \cos\omega t +B \sin\omega t$ to the observations.  The periodogram is written as
\begin{equation}
P=\frac{1}{2} \left\{ \frac{\left[ \sum_{i}x_{i} \cos\omega \left( t_{i}-\tau \right) \right]^{2}}{\sum_{i} \cos^{2}\omega \left (t_{i}-\tau \right) } + \frac{\left[ \sum_{i}x_{i} \sin\omega \left( t_{i}-\tau \right) \right]^{2}}{\sum_{i} \sin^{2}\omega \left( t_{i}-\tau \right)} \right\} \, ,
\label{modified}
\end{equation}
where $\tau$ is given by $\tan\left( 2\omega \tau \right )=\sum_{i} \sin2\omega t_{i}/\sum_{i} \cos2\omega t_{i} $\,.  The periodogram is evaluated for $N_{\nu}$ number of frequencies between the minimum, $\nu_{min} = 1/T$, and the maximum frequencies, $ \nu_{max}=1/(2 \Delta t)$.   The total number of frequencies can be empirically given as $N_{\nu}=n_{0} T \nu_{max} $,  where $n_{0}$ can be chosen in the range of $5-10$ 

In principle, origin of QPOs can be conceived of mainly in three scenarios: supermassive binary black holes (SMBBH) system, accretion disk and jet instabilities.  In SMBBH system, the observed timescales can be interpreted as the Keplerian periods of the secondary black hole around the central black hole and be given by $T=2\pi a^{3/2} (G\ M)^{-1/2}$ with $M$, the total mass be of the system could be order of $\sim 10^{10} M_\odot$, $a$ is the length of the semi-major axis of the elliptic orbit, could represent a distance a few parsecs for a stable configuration. Over the long course of merging galaxies, the dynamical friction present in the system can gradually smooth the elliptical orbits into circular orbits.   Such a binary systems can undergo  orbital decay due to emission of low frequency (a few tens of nano-Hertz) gravitational waves (GW), which could be detected by future GW missions. In SMBBH system the secondary can perturb the accreation of the primary leading to the precession of the disk, which consequently can lead to jet precession.
In the accretion based models.  year time scale periodicity in blazars can be explained in terms of instabilities at the accretion disk. For example, a bright hot-spot could be revolving around the central black hole in a Keplerian period,    Similarly, in the case of globally perturbed thick accretion disks,  the disk can undergoes p-mode oscillations  \citep[e.g see][]{An2013}.   The gravitational field near a fast spinning  supermassive black hole, the frame dragging effect can warp  the inner part of the accretion disk. This might lead to the nodal precession of the tilted plane of the disk better known as the \emph{ Lense-Thirring precession}. 
  The observed quasi-periodic flux modulations can also be linked to relativistic motion of the emission regions along  helical path of the magnetized jets \cite[e.g.,][]{Camenzind92}.  QPOs can also arise due to regular change in the  Doppler factor, that is, changes in either in the angle and/or the boost \cite{Bhatta2018d} .

\section{Conclusion}
To conclude, the main take away of the \gama-ray variability study of blazars can be summarized as the following.
\begin{itemize}
\item{The sample sources are highly variability as shown by the fractional variability}
\item{The \gama-ray flux distribution can be represented  with the log-normal PDF}
\item{the flux distribution exhibits a linear RMS-flux relation}
\item{Blazar PSD is  consistent with flicker noise indicating long memory process at work.}
\item{In some of the sources, year-timescale QPOs are found to be significant.}

\end{itemize}

\section{Acknowledgment}
G.B. acknowledges the financial support by the Narodowe Centrum Nauki (NCN) grant UMO-2017/26/D/ST9/01178.

%
%
%


\begin{thebibliography}{99}
\bibitem{An2013} An, T., Baan, W.~A., Wang, J.-Y., Wang, Y., \& Hong, X.-Y.\ 2013, MNRAS, 434, 3487

\bibitem{Atwood2009}Atwood, W.B., Abdo, A.A. et al., 2009, ApJ, 697, 1071

\bibitem{Bhatta2020a} Bhatta, G. \& Dhital, N.\ 2020, ApJ, 891, 120. doi:10.3847/1538-4357/ab7455

\bibitem{Bhatta2020b} Bhatta, G., P{\'a}nis, R., \& Stuchl{\'\i}k, Z.\ 2020, \apj, 905, 160. doi:10.3847/1538-4357/abc625

\bibitem{Bhatta2019} Bhatta, G.\ 2019, MNRAS, 487, 3990. doi:10.1093/mnras/stz1482 

\bibitem{Bhatta2018d} Bhatta, G.\ 2018, Galaxies, 6, 136

\bibitem{Bhatta2018c}Bhatta, G., Mohorian M., and Bilinsky I.\ 2018  A\&A, 619, A93

\bibitem{Bhatta2018b} Bhatta, G., Stawarz, {\L}., Markowitz, A., et al.\ 2018, ApJ, 866, 132

\bibitem{Bhatta2018a} Bhatta, G., \& Webb, J.\ 2018, Galaxies, 6, 2

\bibitem{Bhatta2017} Bhatta, G.\ 2017, ApJ,, 847, 7

\bibitem{bhatta16c} Bhatta, G., Zola S., Stawarz, {\L}., et al.\ 2016c, ApJ, 832, 47

\bibitem{Bhatta2016b} Bhatta, G., Stawarz, {\L}., Ostrowski, M., et al.\ 2016b, ApJ, 831, 92


\bibitem{Biteau2012} Biteau, J., \& Giebels, B.\ 2012, \aap, 548, A123

\bibitem{Blandford1977} Blandford, R.~D., \& Znajek, R.~L.\ 1977, MNRAS, 179, 433

\bibitem{Blazejowski2000}B{\l}a{\.z}ejowski, M., Sikora, M., Moderski, R., \& Madejski, G.~M.\ 2000, ApJ, 545, 107

\bibitem{Camenzind92}Camenzind M., Krockenberger M., 1992, A\&A, 255, 59

\bibitem[Dermer \& Schlickeiser(1993)]{Dermer1993} Dermer, C.~D., \& Schlickeiser, R.\ 1993, ApJ, 416, 458

\bibitem{Fan2016} Fan, J.~H., Yang, J.~H., Liu, Y., et al.\ 2016, ApJS, 226, 20

\bibitem{Fossati1998} Fossati, G., Maraschi, L., Celotti, A., Comastri, A., \& Ghisellini, G.\ 1998, MNRAS, 299, 433

\bibitem{Ghisellini2017} Ghisellini, G., Righi, C., Costamante, L., \& Tavecchio, F.\ 2017, MNRAS, 469, 255

\bibitem{Ghisellini2011} Ghisellini, G., Tavecchio, F., Foschini, L., \& Ghirlanda, G.\ 2011, MNRAS, 414, 2674

\bibitem{Ghisellini1998} Ghisellini, G., Celotti, A., Fossati, G., et al.\ 1998, MNRAS, 301, 451


\bibitem{Ghisellini2005} Ghisellini, G., Tavecchio, F., \& Chiaberge, M.\ 2005, A\&A, 432, 401. doi:10.1051/0004-6361:20041404

\bibitem{Ghisellini2014} Ghisellini, G., Tavecchio, F., Maraschi, L., et al.\ 2014, \nat, 515, 376

\bibitem{Lyubarskii1997} Lyubarskii, Y.~E.\ 1997, \mnras, 292, 679

\bibitem{Lomb76}Lomb, N. R. 1976, Ap\& SS, 39, 447

\bibitem{Maraschi1992} Maraschi, L., Ghisellini, G., \& Celotti, A.\ 1992, ApJL, 397, L5

\bibitem{Mastichiadis2002} Mastichiadis, A., \& Kirk, J.~G.\ 2002, PASA, 19, 138

\bibitem{2016Galax...4...37M} Marscher, A.\ 2016, Galaxies, 4, 37

\bibitem{Marscher14} Marscher, A.~P.\ 2014, ApJ, 780, 87

\bibitem{Marscher2008} Marscher, A.~P., Jorstad, S.~G., D'Arcangelo, F.~D., et al.\ 2008, Nature, 452, 966

\bibitem{Marscher1992} Marscher, A.~P., Gear, W.~K., \& Travis, J.~P.\ 1992, Variability of Blazars, 85

\bibitem{Mattox1996} Mattox, J.~R., Bertsch, D.~L., Chiang, J., et al.\ 1996, \apj, 461, 396

\bibitem{Press1978} Press, W.~H.\ 1978, Comments on Astrophysics, 7, 103

\bibitem{Scargle82}Scargle, J. D. 1982, ApJ, 263, 835

\bibitem{Sikora2000} Sikora, M. \& Madejski, G.\ 2000, ApJ, 534, 109. doi:10.1086/308756

\bibitem{Sikora1994} Sikora, M., Begelman, M.~C., \& Rees, M.~J.\ 1994, ApJ, 421, 153. doi:10.1086/173633

\bibitem{Vaughan2003}Vaughan, S., Edelson, R., Warwick, R. S., \& Uttley, P., 2003, MNRAS, 345, 1271
....

\end{thebibliography}
\end{document}